\documentclass[%
reprint,
superscriptaddress,
 amsmath,amssymb,
 aps, prl,
]{revtex4-2}

\newcommand{\Otwo}{$\text{O}_{\text{2}}\ $}
\newcommand{\interface}{$\text{Si/SiO}_{\text{2}}\ $}
\newcommand{\SiOtwo}{$\text{SiO}_{\text{2}}\ $}

\usepackage{siunitx}
\usepackage{graphicx}
\usepackage{layouts}
\usepackage{xcolor}

\begin{document}
\vspace{-0.5cm}

\title{%
	Modeling the Initial Stages of Si(100) Thermal Oxidation: An Ab-Initio Approach
	}

\author{Lukas Cvitkovich}
\email[]{cvitkovich$\mid$grasser@iue.tuwien.ac.at}
\affiliation{Institute for Microelectronics, Technische Universität Wien}

\author{Dominic Waldhör}
 \affiliation{Institute for Microelectronics, Technische Universität Wien}

\author{Al-Moatassem El-Sayed}
\affiliation{Institute for Microelectronics, Technische Universität Wien}
\affiliation{Nanolayers Research Computing, Ltd.\\
			Granville Court, Granville Road, London N12 0HL, United Kingdom}%

\author{Markus Jech}%
\affiliation{Institute for Microelectronics, Technische Universität Wien}

\author{Christoph Wilhelmer}
\affiliation{Christian Doppler Laboratory for Single-Defect Spectroscopy in Semiconductor Devices at the Institute for Microelectronics, Technische Universität Wien}

\author{Tibor Grasser}
\affiliation{Institute for Microelectronics, Technische Universität Wien}

\begin{abstract}
    Silicon together with its native oxide \SiOtwo was recognized as an outstanding material system for the semiconductor industry in the 1950s. In state-of-the-art device technology, \SiOtwo is widely used as an insulator in combination with high-$k$ dielectrics, demanding fabrication of ultra-thin interfacial layers.
    The classical standard model derived by Deal and Grove accurately describes the oxidation of Si in a progressed stage, however, strongly underestimates growth rates for thin oxide layers.
	Recent studies report a variety of oxidation mechanisms during the growth of oxide films in the range of \SI{10}{\angstrom} with various details still under debate.
    This paper presents a first-principles based approach to theoretically assess the thermal oxidation process of the technologically relevant Si(100) surface during this initial stage.
    Our investigations range from the chemisorption of single \Otwo molecules onto the $p(2\times2)$ reconstructed Si surface to 
    completely oxidized Si surface layers 
    with a thickness of up to \SI{20}{\angstrom}.
	The initially observed enhanced growth rate is assigned to barrierless \Otwo chemisorption events upon which the oxygen molecules dissociate.
	 We present strong evidence for an immediate amorphization of the oxide layer from the onset of oxidation.
	 Following the oxidation of the first Si layers, direct dissociative adsorptions vanish due to the emergence of an adsorption barrier. 
	 Instead, a slower oxygen incorporation mechanism is mediated by molecular precursor states that spontaneously dissociate after a few ps.
	This process dominates until the surface is saturated with oxygen and separated from the Si substrate by a \SI{5}{\angstrom} transition region.
	Initiating the next stage, the saturated surface becomes inert to any dissociative reactions and enables the diffusion of molecular oxygen to the \interface interface as assumed within the Deal-Grove model.
	Further oxidation of the Si substrate is then provided by \Otwo dissociations at the interface due to the same process responsible for the chemisorption at the surface.
\end{abstract}


\maketitle

\section{INTRODUCTION}
\vspace*{-1cm}
\section{\label{INTRODUCTION}}
A key requirement in micro- and nanotechnology device fabrication is the controlled production of layered material components.
The semiconductor industry relies on silicon and its native oxide \SiOtwo as a prime material system due to their well controlled manufacturing and outstanding interface properties~\cite{Razeghi2009}.
Although pure \SiOtwo is gradually replaced as a gate dielectric by other materials with much higher dielectric constants~\cite{Waldrop2016, 2D2020} -- generally referred to as high-$k$ dielectrics -- an ultra-thin \SiOtwo passivation layer that is grown on the Si substrate before the deposition of the high-$k$ film substantially improves the device performance~\cite{HighK, HighK2, Nitride} and is therefore still mandatory.
Recent fabrication and down-scaling trends shift research interests toward low-temperature chemical-based bottom-up fabrication approaches~\cite{NUR2020, 
HfO2onSi/Ge, lowTOx2007} in which the production of ultra-thin \SiOtwo layers is of crucial importance.
\SiOtwo layers are typically fabricated via thermal oxidation of Si. The physical mechanisms of this process have been investigated for decades, both experimentally and theoretically. 
Today, thicknesses of dielectric a-SiO$_2$ layers demanded for modern semiconductor technologies are in the range of a few nanometers, that is the realm of the initial stage of thermal oxidation. Thus, details of the initial oxidation stage that have not been subject of previous research, which was focused on thicker layers, have become more and more relevant. Problems like the onset of amorphization or the interplay between various oxidation mechanisms are still unresolved.

Among the various theoretical assessments of thermal oxidation, the seminal model developed by Deal and Grove accurately describes the later stages of the silicon oxidation process~\cite{DealGrove}.
The model is based on the well-established concept for Si oxidation that \Otwo molecules enter the \SiOtwo surface layers and diffuse to the \interface interface where they dissociate and individually relax into their respective lattice sites~\cite{Bongiorno2005, Bongiorno2004, Bongiorno2004_2, Pasquarello1998, interface1988, AKIYAMA2005, Gusev1995, Rosencher1979}. However, these assumptions seem to be only valid for a well-progressed oxididation stage (oxide thickness $>\SI{30}{\angstrom}$) as the model predictions strongly underestimate the growth rates for thin oxide films~\cite{Hopper1975, Massoud1985_exp}. 
Massoud \textit{et al.} experimentally determined the oxidation rate for thin layers and extended the Deal-Grove model by adding non-physical exponential terms in order to account for the initially enhanced growth rate~\citep{Massoud1985_exp, Massoud1985_the, Massoud1987}.
Motivated by ab-initio calculations that suggested a strain-driven emission of Si atoms from the Si/SiO$_2$ interface, Kageshima \textit{et al.} proposed a model that includes oxidation reactions within the oxide layer~\cite{Kageshima1998, Kageshima1999}. Under these assumptions, almost perfect agreement with experimental data could be achieved for oxide thicknesses $>\SI{5}{\angstrom}$.
Recent ab-initio calculations showed that the initial oxidation of a clean and reconstructed Si surface is based on chemisorption events resulting in the dissociation of the adsorbing \Otwo molecule directly at the surface~\cite{InitialAdsorp, InitialAdsorp2, InitialAdsorp3}. 
This was confirmed experimentally in recent electron microscopy and photoemission studies~\cite{Liao2006, expInitialAdsorp16, STM2020}.
In addition, metastable molecular surface states were observed by means of scanning electron microscopy and electron spectroscopy techniques on thin oxide layers~\cite{Precursor, Precursor2, STM2020, Ferguson1999, Morgen1989}. 
At low temperatures, these states precede dissociation events, however, vanish upon annealing. Hence, they are merely intermediate states toward the dissociative surface reaction.
Both adsorption types, direct and molecular precursor mediated, were observed in molecular beam experiments~\cite{MB1999}. According to this study, incident \Otwo molecules of high kinetic energy ($> \SI{1}{\electronvolt}$) tend to dissociate directly, whereas the molecular precursor states originate from trapping-mediated adsorption of \Otwo with low kinetic energy ($< \SI{0.2}{\electronvolt}$).
Upon dissociation, the O atoms moved into backbonds of Si surface atoms, as correctly predicted by the above mentioned theoretical studies~\cite{InitialAdsorp, InitialAdsorp2, InitialAdsorp3}.

Combing these results, two major kinetic mechanisms seem to be at work: surface reactions dominate the initial stage, while the diffusion of \Otwo becomes more important as the oxidation continues. 
Based on this concept, a theoretical growth model that considers dissociative chemisorption at the surface has been formulated and the resulting growth rates agree with experimental observations~\cite{NewModel95}.
Further experiments indeed support the concept of two distinguishable oxidation stages.
As reported in a photoelectron spectroscopy study~\cite{Hoshino2001}, the initial stage at low \Otwo exposure of up to 10~L (1 L: $10^{-6}$ Torr s) featured relatively rapid oxidation.
Maintaining the supply of O$_2$, the oxygen surface coverage saturated and the oxidation rate was reduced as the system gradually transitioned into the Deal-Grove regime. 
According to transmission electron microscopy (TEM) experiments on Si nanoparticles, diffusion of \Otwo molecules becomes important after formation of a \SI{5}{\angstrom} thick oxide layer~\cite{Liao2006}.

Besides establishing the fundamental processes of oxidation, the origin of amorphization is still a controversial issue in literature.
Although some studies predicted the initial growth of a crystalline Si oxide monolayers ~\cite{Fuchs2005, Anne}, an increase in the surface roughness during the initial oxidation was observed by atomic force microscope measurements~\cite{AFM2009} indicating an amorphous oxide growth.
More recent TEM measurements support this idea and reported evidence that amorphous oxides were obtained by rapid thermal oxidation (RTO)~\cite{MUR2001, Liao2006}. 

In this work the initial stage of thermal oxidation of Si was studied up to an oxide thickness of \SI{20}{\angstrom} in order to investigate the growth and amorphization of ultra-thin Si oxide layers as demanded by novel devices operating on the nano scale.
Preliminary results were already published in our recent work~\cite{CVITKO2021}. However, the present study is far beyond and provides a comprehensive bottom-up approach focused on the thermal oxidation and its intricacies associated with its various stages.
For the first time, the complete growth of an ultra-thin \SiOtwo layer was realistically modeled within dynamic simulations in full agreement with all experimental and theoretical knowledge that has been gained on this subject until now.
Earlier ab-initio studies were restricted to static calculations or used vicarious crystalline SiO$_2$ surfaces because amorphous materials could not be treated within ab-initio calculations~\cite{Kageshima1998, Kageshima1999, Fuchs2005, Anne, InitialAdsorp2, InitialAdsorp3}.
The above mentioned TEM experiments only studied oxide surfaces with more than five monolayers (ML) of \SiOtwo (from crystalline SiO$_2$:~1~ML:~\SI{6.78e14}{atoms/cm^2}~\cite{Hoshino2001}). In contrast, our simulations go beyond these measurements and offer detailed insight into the oxidation process from the onset of oxidation.
The thermal oxidation was modeled by the subsequent adsorption of \Otwo molecules within ab-initio molecular dynamics (AIMD) simulations in conjunction with density functional theory (DFT).
Energy barriers for the migration of oxygen in the oxide layers were obtained by nudged elastic band (NEB) calculations.
The initial oxidation process featured the highest oxidation rate enabled by spontaneous surface reactions.
An immediate amorphization was indicated 
by arbitrarily incorporated oxygen along many possible adsorption trajectories. 
The oxidation led to
the spontaneous formation of SiO$_4$ tetrahedrons, the characteristic structural elements of a-SiO$_2$. 
With increasing O coverage, we observed that O-coordinated Si atoms were less susceptible for \Otwo chemisorptions.
Hence, repulsion of the incident \Otwo became more probable and the oxidation rate decreased gradually as the surface was oxidized. 
In this stage, direct dissociation ($<\SI{0.5}{\pico\second}$) was only possible for \Otwo molecules of high kinetic energy
while the molecular precursor states were typically observed for low energy molecules leading to a further decrease in the oxidation rate.
The subsequent transition to the diffusion limited regime and the associated migration of O$_2$ was investigated on a $\SI{20}{\angstrom}$ thick surface layer of a-SiO$_2$. \Otwo diffusion became the dominate mechanism as soon as a sufficiently thick surface layer was saturated with O. The saturation was consistent with the formation of a-SiO$_2$ and indicated the transition into the Deal-Grove regime. A spontaneous dissociation, similar to the surface reaction, was then observed at the \interface interface. 
An overview of the mechanisms and their onset during the oxidation process is given in Fig.~\ref{scheme}.
%

\section{METHODOLOGY}
\begin{figure*}[htbp]
	\centerline{\includegraphics[width=0.9\linewidth]{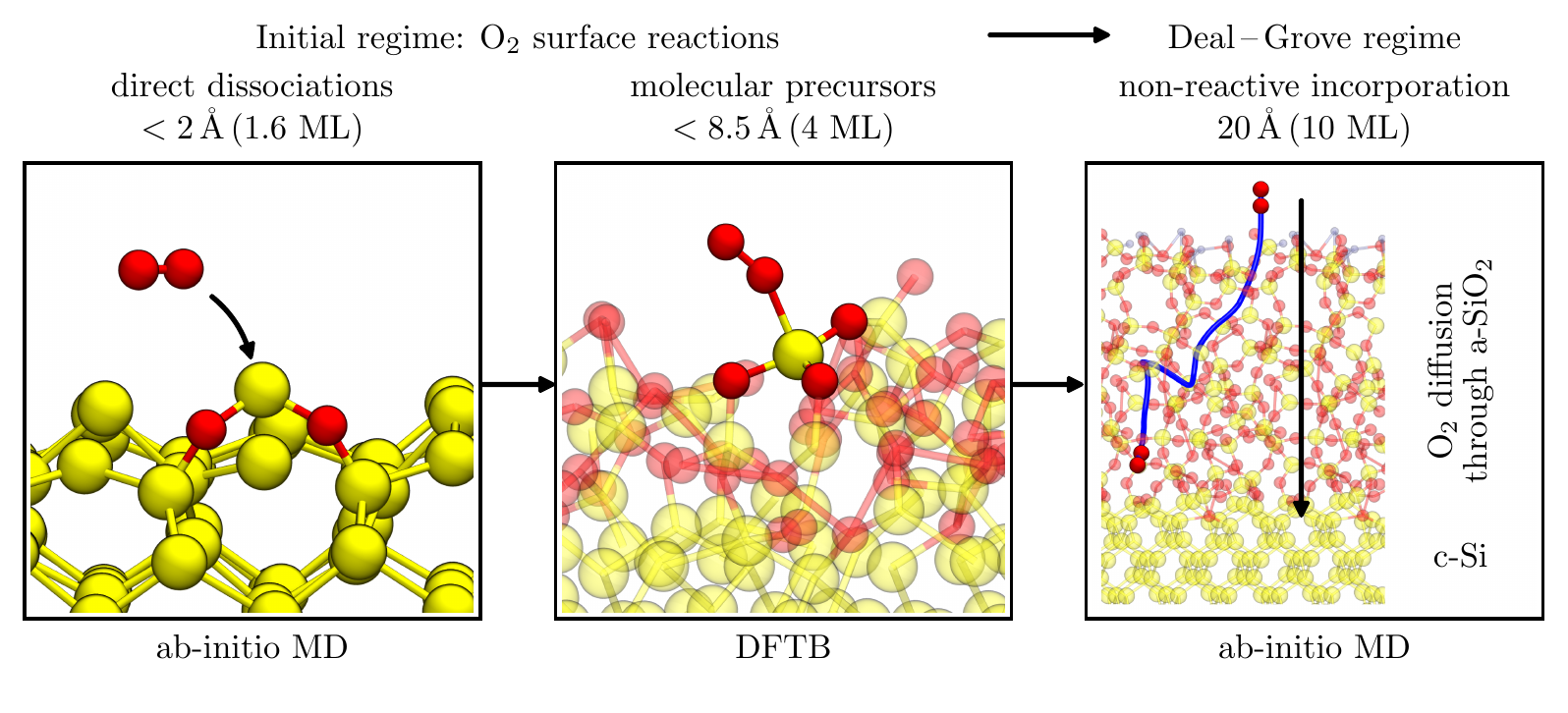}}
    \caption{Simulation scheme: We modeled the growth of \SiOtwo starting from a clean Si(100) surface. In the initial stage, spontaneous and dissociative adsorptions of \Otwo molecules were observed in DFT-based ab-initio MD (AIMD) calculations. These surface reactions are responsible for the increased growth rate compared to the Deal-Grove regime and lead to an immediate amorphization of the surface oxide.
    After oxidation of the first layer (oxygen coverage of 1~ML), an adsorption barrier formed and required imposing initial velocities onto the \Otwo molecules. As soon as the oxygen coverage reached 1.6~ML, the predominant adsorption mechanism changed to a slower process exhibiting molecular \Otwo precursors that dissociated after a few~ps. In order to account for the longer simulation times needed, we employed DFTB for the further oxidation of the Si slab until a \SI{8.5}{\angstrom} oxide layer formed which showed structural characteristics of bulk a-SiO$_2$.
Eventually, the saturated surface became inert to any surface reactions and \Otwo diffusion to the \interface interface -- the fingerprint of Deal-Grove oxidation -- set in as observed on a \SI{20}{\angstrom} thick oxide.
}
    \label{scheme}
\end{figure*}

Our calculations were mainly conducted at the ab-initio level utilizing density functional theory (DFT). In order to study oxidation mechanisms beyond the initial steps of O adsorption and dissociation, further investigations inevitably had to be carried out on larger model systems. Especially for DFT calculations, the computationally feasible simulation time was limited to a few ps and could be even lower for structures with an increased density of crystallographic defects. 
To be able to further extend our data set,
we additionally used density functional based tight binding (DFTB) in conjunction with a Slater-Koster parameter set designed for Si surfaces and interfaces with SiO$_2$~\cite{DFTB+, SKparameter}.
Initial benchmarks showed that both methods yield comparable results.
The details of the utilized methods and their applications within this work (cf.~Fig.~\ref{scheme})
are summarized below.

\subsection{DFT setup}
All density functional theory calculations were carried out using the CP2K package~\cite{Quickstep}, a code that uses the mixed Gaussian and plane waves approach (GPW). We used a double-$\zeta$ Gaussian basis set that was optimized for condensed-phase systems consisting of Si, O, and H and the well-established Goedecker-Teter-Hutter (GTH) pseudopotentials 
\cite{BasisSet, Goedecker1996}. 
The electron density was expanded using a plane-wave basis with a cutoff of \SI{650}{Ry}.
The exchange-correlation energy was obtained
by means of the semilocal 
generalized gradient approximation (GGA) functional PBE. 
Due to its efficiency, the orbital transformation (OT) method was used by default to find the ground-state electronic structure~\cite{OT}.
However, within dynamic AIMD calculations, where several O atoms and potentially multiple unsatured bonds were involved,  Broyden's method proved to be more stable and was used instead~\cite{Broyden1970}.
Atomic relaxations were carried out using the Broyden-Fletcher-Goldfarb-Shanno  (BFGS)  algorithm~\cite{Goldfarb1970}
with a force convergence criterion of 
$\SI[per-mode=repeated-symbol]{15}{\milli\electronvolt\per\angstrom}$.
Within the AIMD
simulations, the total energy was conserved (microcanonical or NVE ensemble) and the total spin was 
restricted to $S=0$. The minimal energy barrier between two configurations was calculated using the climbing-image NEB (CI-NEB) 
method~\cite{CI-NEB1,CI-NEB2} with a spring constant of
$k=\SI[per-mode=repeated-symbol]{8.2}{\electronvolt\per\angstrom\squared}$.

\subsection{DFTB setup}
The DFTB method uses an expansion of the total energy
of DFT 
with respect to the charge density~\citep{DFTB}. Hamiltonian matrix elements and overlap integrals are approximated by interpolations between two-atom calculations obtained from DFT. These approximations reduce the computational costs drastically. 
Our DFTB simulations were carried out using the DFTB+ package~\cite{DFTB+}, employing the
Slater-Koster parameter set 
pbc-0-3~\cite{SKparameter}, which is suitable for solids and surfaces of \hbox{Si-O-H} systems.
We considered atomic basis functions up to the s-, p- and d-shells for H, O, and Si, respectively.
For geometry optimizations again the BFGS algorithm implemented in the DFTB+ package was utilized.
The DFTB approach as well as the pbc-0-3 parameter set~\cite{SKparameter} are well justified as reported by a number of previous studies~\cite{ZHENG2005, Capel2015, Tingting2014}.
The agreement between AIMD and DFTB calculations within the scope of the present paper was ensured by a number of test calculations, e.g. the adsorption of \Otwo molecules showed the same charge transfer of $-e$ as well as comparable reaction kinetics and adsorption configurations. Furthermore, recalculation of an AIMD trajectory reassuringly gave very similar energies. The comparative calculations are given in the supplementary material.

\subsection{Preparation of atomic structures}
The starting point of our investigations was a $4\times4\times12$ reconstructed Si surface structure. 
A cleaved Si surface leads to undercoordinated Si atoms at the surface that reconstruct by forming alternating rows of tilted dimers to minimize its energy. This reconstruction reduces the number of dangling bonds on the surface via electron transfer from the lower Si dimer atom to the upper one~\cite{dimers}. In the present structure eight dimer pairs formed within the simulation cell. The dimers were aligned in rows that build terraces on the surface separated by cavities, the so-called channels. 
The dangling bonds at the bottom of the structure were passivated with hydrogen. The bottom Si layer and the passivating H atoms were fixed in AIMD runs to resemble a bulk like structure.
After reconstruction of the surface, optimizing the cell including the ionic cores in the lateral directions within DFT resulted in cell dimensions of $a=b=\SI{15.523}{\angstrom}$. The cell size in the $c$-direction was set to $c=\SI{37.22}{\angstrom}$ leaving a vacuum of \SI{20}{\angstrom} above the slab. 
The thermal oxidation of Si was studied on this model by means of AIMD 
until 1.6~ML of O were adsorbed.
Thereafter, we employed DFTB to model the thermal oxidation up to an oxide thickness of \SI{8.5}{\angstrom}.
Investigations of \Otwo migration through the oxide were carried out on a thicker oxide model which was obtained by a melt and quench procedure~\cite{Jech, Tassem2014, MeltQuench2004, MeltQuench2012, MeltQuench2015} using classical molecular dynamics. Details of this procedure are given in Ref.~\cite{Jech}. The in this way obtained Si/SiO$_2$/Si structure was split at the oxide to generate a surface which was passivated with H and relaxed within DFT. The finished surface model consisted of a $3\times3\times12$ Si substrate with roughly \SI{20}{\angstrom} of \SiOtwo on top, adding up to a total of 290 atoms.

\section{Results and Discussion}
\section{\label{RESULTS}}
A detailed picture of oxygen incorporation and amorphization at the initial stage of thermal oxidation of Si could be established from the outcome of our simulations.
Our results explain the experimentally observed transition from a fast to a slow oxidation regime~\cite{Hoshino2001, Liao2006} by various oxygen incorporation mechanisms that supersede each other during the oxidation of only a few layers of Si. The key findings presented in this section are:
\begin{itemize}
  \item Spontaneous surface reactions upon which the involved \Otwo molecules dissociated.
  \item Immediate amorphization of the oxide layer.
  \item Molecular precursor states provided for a slower oxidation rate in a more progressed stage of oxidation.
  \item Oxygen surface saturation marked the beginning of \Otwo diffusion through the oxide.
  \item Dissociation at the \interface interface due to the same charge transfer process that characterized the initial surface reactions.
\end{itemize}
\subsection{Dynamic simulations of thermal surface oxidation}
Thermal oxidation, typically in the range of \SI{1000}{\kelvin}, is a highly dynamic and complex process. Previous theoretical attempts to mimic the initial stage, however, mainly relied on static calculations leading to crystalline oxide monolayers~\cite{Fuchs2005, Anne}, as is also presented in the supplementary material. Therefore, accounting for lattice dynamics is a mandatory prerequisite for credible calculations. Our simulations clearly show that the various adsorption trajectories and sites strongly depend on these random movements of the involved atoms.
This results in a stochastic adsorption process which ultimately prevents the growth of an ordered oxide. 
Our investigations on the DFT level started from the Si(100) surface with a $p(2\times2)$ reconstruction and ranged up to the adsorption of 13 \Otwo molecules, corresponding to a coverage of 1.6~ML in our structures, offering insight into the amorphization of the first oxide layer.
In order to study the further progression of oxidation, we employed DFTB in the following until a \SI{8.5}{\angstrom} (4~ML) thick oxide layer was obtained.		

\subsubsection{
\Otwo adsorption onto the clean Si surface}
\begin{figure}[htbp]
	\centerline{\includegraphics[width=\linewidth]{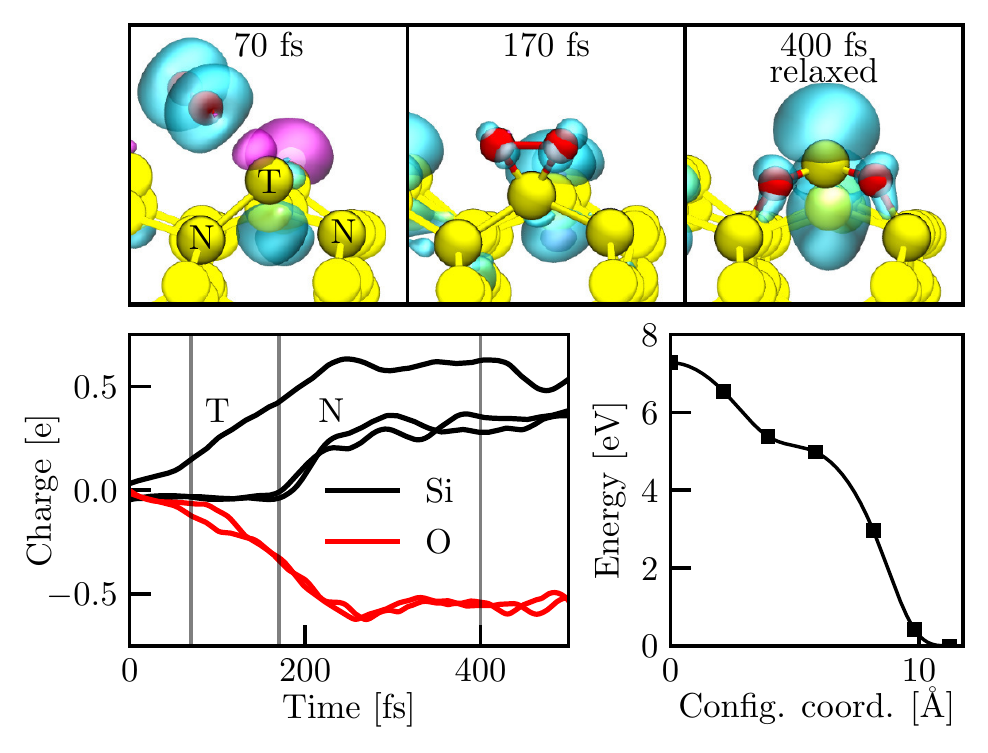}}
    \caption{The dissociative chemisorption of \Otwo (red atoms) onto a clean and reconstructed Si(100) surface (yellow atoms). 
    Snapshots of the AIMD simulation are given in the upper panel. The top right panel shows the geometry optimized structure after dissociation. The spin density for values $\pm\SI{0.002}{e\per\bohr^3}$ is depicted by cyan and magenta isosurfaces, respectively.
    The Mulliken charges of the oxygen atoms, the top dimer Si atom (T), and the neighboring Si atoms (N) 
    during the first \SI{500}{\femto\second}
    of the chemisorption are shown in the lower left panel.
    A barrierless adsorption and dissociation is indicated by a NEB calculation along the adsorption trajectory in the lower right panel.
    }
    \label{charge_transfer}
\end{figure}

Our investigations started with a single \Otwo molecule that was placed \SI{2.5}{\angstrom} above the reconstructed Si(100) surface. 
An ensemble of randomly generated and Maxwell-Boltzmann (MB) distributed velocities, scaled to match the specified temperature $T=\SI{1000}{\kelvin}$, was assigned to all atoms and a spin-restricted ($S=0$) AIMD simulation with a simulation time of \SI{3}{\pico\second} and a step size of \SI{0.5}{\femto\second} was performed. Subsequently, the structure was relaxed within DFT. 
In this vein, several simulations with various starting configurations yielded consistent results. 
Fig.~\ref{charge_transfer} shows snapshots of one representative \Otwo adsorption and dissociation event together with the associated charges obtained by a Mulliken charge analysis. The \Otwo molecule spontaneously moved toward an upper Si dimer atom that was charged positively as the \Otwo molecule approached. 
After \SI{170}{\femto\second} the molecule was centered above the Si dimer atom and started to dissociate. Within the next \SI{80}{\femto\second}, the dissociation process was completed and the charges remained constant for the rest of the simulation. The two oxygen atoms relaxed into the backbonds of the upper Si dimer. In total, a charge of roughly $-e$ was transferred from the surrounding Si surface atoms to the \Otwo molecule. 
In a simple molecular orbital picture, the donated electron occupied the antibonding $\pi^*$ orbital of the \Otwo molecule, triggering its dissociation.
Note that the gradual increase of charge depicted in Fig.~\ref{charge_transfer} resulted from the adiabatic representation within DFT. In reality, the charge would be transferred much faster in a non-adiabatic process.
Regardless, a spontaneous dissociative surface reaction was indicated by the very rapid reaction with the Si surface that was found to be independent from the starting configuration.
This result was further supported by a series of static calculations. Placing \Otwo molecules at random positions above the clean and reconfigurated Si(100) surface and relaxing the structure within DFT led to the dissociation of the molecule.
Furthermore, a barrierless dissociative oxidation was also reported in \cite{InitialAdsorp2}.
Compared to the pre-adsorbed state, the intact \Otwo molecule above the pristine Si slab, the dissociated and geometry optimized structure was \SI{7.28}{\electronvolt} lower in energy.
As shown in the last snapshot in Fig.~\ref{charge_transfer}, the Si atom on top was still undercoordinated resulting in a surface dangling bond. This exact configuration was already observed experimentally by STM images of a sparsely oxidized Si surface~\cite{STM2020}.

\subsubsection{Initial amorphization}
Crystalline oxide structures on Si surfaces have been investigated thoroughly within theoretical studies~\cite{Fuchs2005, Anne, Kageshima1998} and also within this work, see the supplementary material.
However, the fact that thermally grown \SiOtwo is amorphous was established decades ago and more recent studies reported evidence that this is also true for thin films of \SIrange{10}{50}{\angstrom} obtained by RTO, as reported by various TEM studies~\cite{MUR2001, Liao2006}. 
Despite these results, it is still unclear if the very first oxide layer is already amorphous.
Earlier ab-initio studies were restricted to static simulations~\cite{Fuchs2005, Anne, Kageshima1998} while experimental studies have not looked at single oxide monolayers~\cite{AFM2009, MUR2001, Liao2006}. Within our dynamic simulations, strong evidence for immediate amorphization was found as presented in the following.

We modeled the thermal oxidation process by a series of AIMD simulations at \SI{1000}{\kelvin} in which \Otwo molecules were consecutively introduced above the reconfigurated Si(100) surface as shown in Fig.~\ref{charge_transfer}. 
The individual initial position was assigned randomly within a distance of \SIrange{2}{3}{\angstrom} above the top Si atom. As for the adsorption in Fig.~\ref{charge_transfer}, the velocity was chosen randomly from a MB distribution at $T=1000$~K. 
The axis of the \Otwo molecule was aligned with the surface in order to allow for effective interaction due to larger spatial overlap between the surface dangling bonds and the oxygen $\pi^{*}$ orbital. The simulation time for each adsorption event was set to \SI{1}{\pico\second}. After this time, another \Otwo molecule was introduced similarly to the previous one, while all other atoms continued to move according to their current velocity.
A Mulliken charge analysis showed that the fundamental oxidation mechanism -- a dissociative chemisorption via a charge transfer similar to the reaction shown in Fig.~\ref{charge_transfer} -- underlay every single adsorption during the initial surface oxidation. 
However, adsorption became less probable as the oxidation progressed. Only the first six adsorptions, corresponding to an oxygen coverage of 0.7~ML in our structures, happened spontaneously.
After that, the \Otwo molecules were occasionally repelled from already oxidized Si atoms. In this case, the run was discarded.
After 9 successful adsorptions (or 1.1~ML), a spontaneous adsorption could not be observed within 10 runs.
We interpreted the decrease in adsorption probability as the emergence and subsequent increase of an adsorption barrier with proceeding oxidation.
In order to overcome the formed adsorption barrier and to study further oxidation within a reasonable time frame, we imposed an initial velocity of \SI{1000}{m/s} toward the surface for the \Otwo molecule from this point onward. 
This value roughly complies with the average velocity of $\overline{v}=\SI{810}{\meter/\second}$ according to the MB distribution of non interacting \Otwo molecules at $T=\SI{1000}{\kelvin}$.
Adsorption of \Otwo -- a process that happened spontaneously onto the clean surface within static and dynamic calculations -- was inhibited by the oxide formation on the surface. 
This behavior qualitatively explained the decrease of the oxidation rate measured in the initial stage of oxidation~\cite{Hoshino2001} and gave rise to the non-reactive \Otwo diffusion process later on~\citep{DealGrove}.

The immediate amorphization of the oxide layer is enabled by a stochastic adsorption process in which the adsorption trajectory of each \Otwo molecule depends strongly on its initial position and velocity and even on fluctuations of the surface lattice due to thermal vibrations. The stochastic character of the oxidation process is indicated by many different adsorption trajectories featuring comparable energy gains and strongly varying final configurations. Representative adsorption trajectories of three successive runs are given in the supplementary material.

\begin{figure}[htbp]
	\centerline{\includegraphics[width=\linewidth]{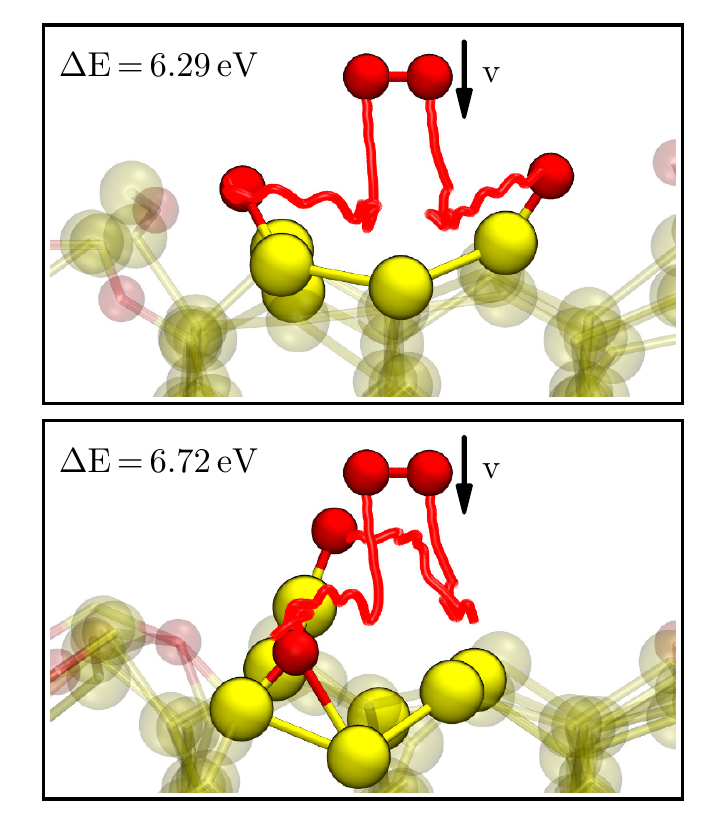}}
    \caption{Two examples of O$_2$ adsorption trajectories of \Otwo adsorptions with the same initial position and velocity of the \Otwo molecule but different velocities for the surface atoms. The velocities of the lattice atoms were randomly sampled from a Maxwell-Boltzmann distribution at 1000~K. The snapshots show the final positions after a simulation time of 450~fs superimposed with the starting configuration. Also shown are the trajectories of the adsorbing O atoms (red lines). The energy gains $\Delta$E were calculated for geometry optimized structures.}
    \label{add_to_clean}
\end{figure}

The adsorption trajectories were very sensitive to changes of the environment as investigated by two comparative AIMD runs. As shown in Fig.~\ref{add_to_clean}, the adsorption of an \Otwo molecule was sampled for two different velocity distributions (taken from a MB distribution as described above) of the surface lattice. The initial position and velocity of the \Otwo molecule was identical for both runs. 
The resulting trajectories are completely different although the energy gain of around 6.5~eV is very similar. This implies that even weak perturbations alter the resulting structure substantially. Note that the respective trajectories are not important and would look different for any ensemble of initial velocities. Hence, Fig.~\ref{add_to_clean} illustrates solely the large effect of slight changes in initial conditions.
The intrinsic high degree of randomness during the oxidation provides strong evidence for the immediate amorphization of the oxide layer.

\subsubsection{Molecular precursors}
Sampling slower surface reactions and more spontaneous adsorptions in which the adsorption barrier was overcome, requires a longer simulation time together with an increased number of calculations. Furthermore, extending our investigations in the direction of the formation of a-\SiOtwo required a larger amount of oxygen. Hence, we conducted these calculations within DFTB. 
The molecules were placed about 2 to \SI{3}{\angstrom} above a surface Si atom before MD simulations of \SI{10}{\pico\second} with a step size of \SI{0.5}{\femto\second} were conducted. 
The first simulation started with random initial velocities (MB distribution at $T=1000$~K) for all surface atoms that were passed on to the subsequent run. 
The \Otwo velocities were sampled from the same distribution though restricted to the negative $z$-direction.
If a molecule was repelled from the surface, it was removed from the simulation cell at the end of the run.

\begin{figure}[htbp]
	\centerline{\includegraphics[width=\linewidth]{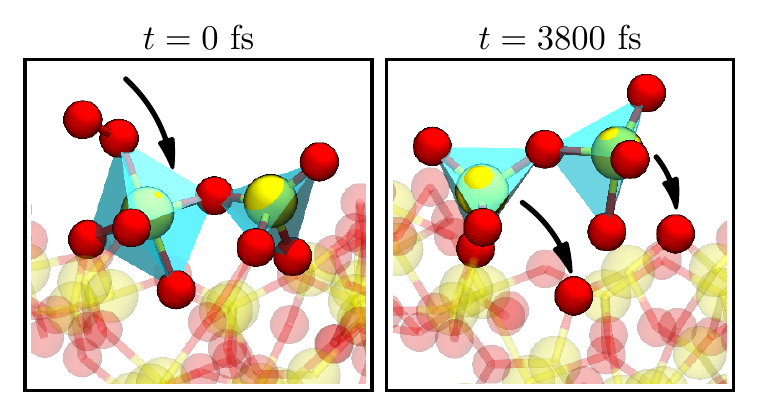}}
    \caption{Molecular precursor states. The \Otwo molecule adsorbs to a already fully oxidized Si atom (the SiO$_4$ tetrahedron is indicated by blue planes) and stays in the vicinity of this atom for around \SI{4}{\pico\second} within our simulations until it finally dissociates. Two O atoms migrate to lower Si atoms adjusting for the incorporation of the dissociated O atoms, as indicated by the black arrows.}
    \label{MolecularPrecursor}
\end{figure}

After the first Si layer was oxidized, spontaneous dissociative adsorptions could hardly be observed. Instead, the \Otwo molecules adsorbed via a metastable molecular precursor state. Dissociation events of these molecules could be sampled within the time-extended DFTB simulations, as shown in Fig.~\ref{MolecularPrecursor}. Again, the dissociation was induced by a charge transfer. The surface was subjected to a considerable reconfiguration, mainly associated with the breakage of multiple \hbox{Si-O} bonds adjust for the dissociation and incorporation in a SiO$_4$ tetrahedron. 
Such precursor states block further adsorption of oxygen molecules and thus effectively lower the rate of oxidation. As mentioned above, we also observed the direct adsorption and dissociation mechanism, provided the O$_2$ molecules possess a sufficiently high kinetic energy, see the supplemental material. However, this process required \Otwo velocities from the top one percent of the MB distribution and was thus ranked negligible in the framework of this work. 
However, both mechanisms involve the breaking of bonds between already dissociated O atoms and neighboring Si which is only feasible if the O atoms find new positions, i.e. can bind to Si atoms below that are not fully oxidized yet. Thus, the rate for direct dissociative oxidation gradually decreases as the oxidation advances and eventually can be considered negligible once a few layers of Si are oxidized. The precursor-mediated process with a reduced oxidation rate is dominant in this regime. 

\subsection{Surface saturation}
\begin{figure}[htbp]
	\centerline{\includegraphics[width=.9\linewidth]{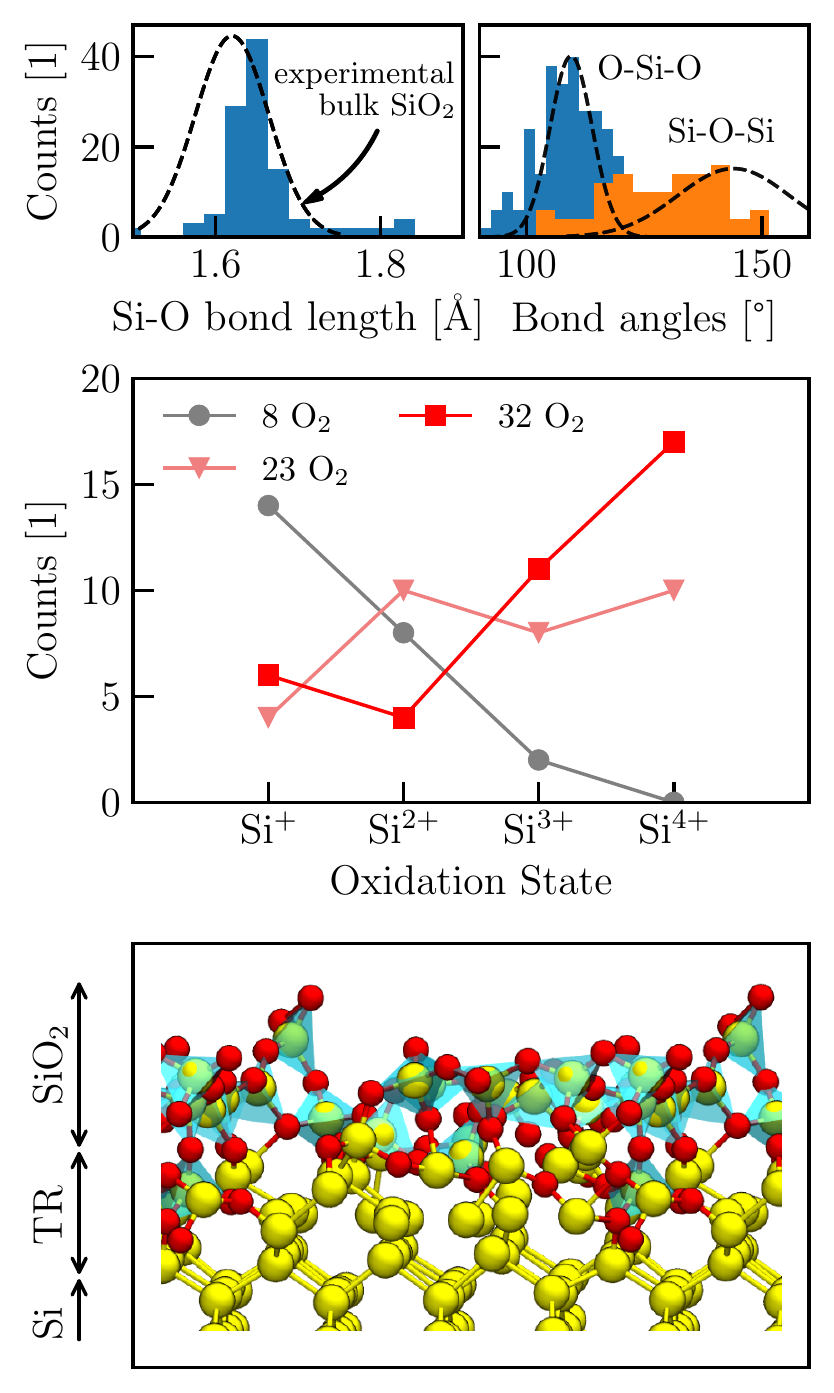}}
    \caption{Characteristics of the \SiOtwo growth obtained by dynamic simulations within DFTB. Distributions of the bond lengths and angles and the comparison to experimental values of bulk \SiOtwo ~\cite{Diebold1999, Mozzi1969} (dashed lines) are shown in the top panel. The prevalence of different Si oxidation states as obtained by a Mulliken charge analysis is shown in the middle panel. As oxidation continues, the amount of Si$^{4+}$ (fully oxidized Si atoms that are incorporated in SiO$_4$ tetrahedrons) increases whereas the amount of Si$^{+}$ converges to a lower value that represents states in the transition region (TR) between the oxide and the crystalline Si substrate. The final and geometry optimized structure is shown in the lower panel. SiO$_4$ tetrahedrons on the immediate surface are indicated by blue planes.}
    \label{SiO2_growth}
\end{figure}

The oxide layer reached a total thickness of \SI{8.5}{\angstrom} after the adsorption of 32 \Otwo molecules or 4~ML. As presented in the following section, the oxide could be subdivided into a \SI{3.5}{\angstrom} thick layer of a-\SiOtwo on top of a \SI{5}{\angstrom} transition region.
In this stage, all Si atoms on the surface were fully O coordinated and incorporated in a spontaneously formed SiO$_4$ tetrahedron, see Fig.~\ref{MolecularPrecursor} and \ref{SiO2_growth}. As a result, the surface structure was highly unordered.
A spontaneous adsorption of \Otwo molecules onto this surface could not be observed, as an additional molecule was repelled from 25 random positions.
The chemical composition of the a-\SiOtwo surface layer was SiO$_{1.95}$.
A structural analysis of this layer showed that the bond lengths and angles are already comparable to the experimentally obtained values for bulk a-\SiOtwo~\cite{Diebold1999, Mozzi1969}: the average \hbox{Si-O} distance was \SI{1.66}{\angstrom} and the binding angles showed an average of \SI{108}{\degree} for O-\hbox{Si-O} and \SI{132.5}{\degree} for \hbox{Si-O-Si}, see Fig~\ref{SiO2_growth}. The slightly larger bond lengths (compared to \SI{1.62}{\angstrom} of bulk SiO$_2$) could be assigned to surface effects. The \hbox{O-Si-O} angles matched the perfect tetrahedral bond angle of \SI{109.47}{\degree}. 
The \hbox{Si–O–Si} angle averaged at \SI{132.5}{\degree}, which agreed well with the conclusions reached in Ref.~\cite{Hirose1999}, that the
angle is reduced to \SI{135}{\degree} in thin films compared to \SI{148}{\degree} in bulk structures.
Furthermore, the \hbox{Si-O-Si} angles were broadly distributed between \SIrange{100}{160}{\degree} which indicated a vitreous (amorphous) form  of silica.
The density of the \SiOtwo layer was roughly \SI{2.33}{\gram/\centi\meter\cubed} in agreement with measured values~\cite{Diebold1999}. The density of the transition layer directly at the \interface interface was slightly increased to values up to \SI{2.7}{\gram/\centi\meter\cubed} which agrees with an intrinsic compressive stress as reported in~\cite{Stress1979, Stress1987}.
A Mulliken charge analysis allowed us to infer the oxidation state of each Si. Due to the strong electronegativity of O, a \hbox{Si-O} bond is represented by an increase of the Si's associated Mulliken charge.
As determined by a reference calculation in defect-free a-SiO$_2$, the fully oxidized state Si$^{4+}$ corresponded to an increase of roughly $1e$ in our DFT setup, yielding $\sim0.25e$ of excess charge on the Si ion per \hbox{Si-O} bond.
The prevalence of different Si oxidation states as determined by this relation during the thermal oxidation is shown in the middle panel in Fig.~\ref{SiO2_growth}. 
For a sparsely oxidized surface, the O was distributed evenly on the Si atoms as shown by the large number of Si$^+$.
In later stages however, most Si were fully O coordinated with only a few partially O-coordinated Si atoms that were located at the interface to the Si substrate.
Si$^{4+}$ atoms were fully coordinated by O and incorporated in a SiO$_4$ tetrahedron. We observed that the transition into the \Otwo diffusion regime (Deal-Grove regime) happened as soon as a sufficiently thick surface layer was fully oxidized. In this stage, the \Otwo could not chemically react with the surface anymore and thus diffusion of \Otwo through the oxide set in.

\subsection{Si/SiO$_x$ interface}
Within computational material modeling, the construction of credible interface structures between amorphous oxides and crystalline substrates is a challenging task~\cite{LING2013, Khalilov2011}. Typically, computationally modeled \interface interfaces are created using a melt and quench procedure~\citep{Jech, Tassem2014, MeltQuench2004, MeltQuench2012, MeltQuench2015}. 
In this approach, atomistic structures are melted at temperatures of up to \SI{7000}{\kelvin} within simulation times of tens of picoseconds while ensuring that the silicon maintains its crystalline structure and the oxygen is confined to a certain region.
In contrast, by consecutively adsorbing \Otwo molecules onto the Si surface in dynamic simulations, we naturally obtained a credible interface between the amorphous oxide and the crystalline Si substrate.
When using the melt and quench method, a very low defect density can be accomplished. Even completely defect-free configurations can be generated~\cite{Jech, Tassem2014}. The stepwise oxidation on the other hand gave interfaces that show larger stresses and more defects (e.g. over/undercoordinated atoms).
However, a reliable quantitative comparison of defect densities could not be given within the scope of this work since the simulation time of our AIMD/DFTB calculations lies still far away from those reached with classical MDs and thus did not allow for a reconfiguration of the amorphous surface layer.

\begin{figure}[htbp]
	\centerline{\includegraphics[width=\linewidth]{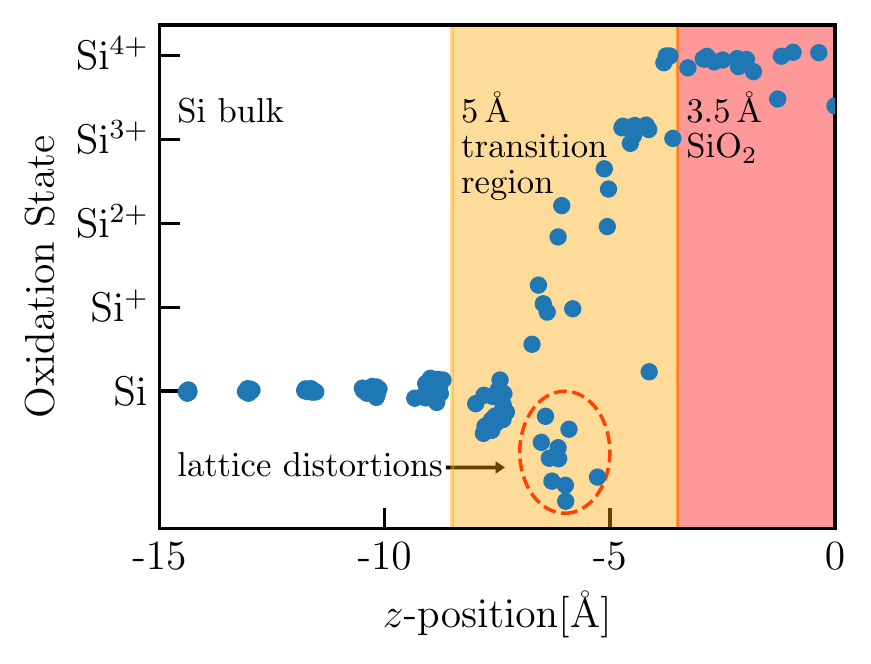}}
    \caption{Oxidation states as obtained by a Mulliken charge analysis of Si atoms vs. their respective $z$-position at an oxygen coverage of 4~ML. Amorphous SiO$_2$ is indicated by fully oxidized Si atoms (oxidation state +4). The oxidation states in the roughly \SI{5}{\angstrom} thick transition region can only be assigned approximately and are blurred by lattice distortions.}
    \label{q_vs_z}
\end{figure}

In order to investigate oxidation states at the interface, we plotted the Mulliken charges of the Si atoms vs.~the corresponding $z$-position, as shown in Fig.~\ref{q_vs_z}. Note that the oxidation state ${+}1$ that is associated with one \hbox{Si-O} bond corresponded to a charge transfer of $0.25e$ in the Mulliken charge analysis. 
All Si atoms in the surface \SiOtwo layer were fully oxidized and their associated Mulliken charge was increased by $e$. The transition region was about \SI{5}{\angstrom} thick and could be identified via intermediate charge states of the silicon atoms. The linear transition of associated charges in the transition region was also found in structures obtained by the melt and quench method~\cite{Jech}.
In agreement with previous studies that reported an intrinsic stress around the interface~\cite{Stress1979, Stress1987}, we observed stretched bonds near the oxidation front that exceeded typical Si-Si binding lengths by about 4\% even for correctly coordinated Si atoms. Ultimately, this led to assymetric Si tetrahedrons with asymmetric charge distributions which altered the assigned Mulliken charge.
Hence, only a rough assignment of oxidation states in the transition region was attainable.
In particular, the decrease of the charge of some Si atoms in the transition region was considered to be an artifact of these lattice distortions.
As discussed above, O-Si-O angles in the oxide matched the perfect tetrahedral bond angle of 109.47°, indicating that these tetrahedrons were practically not distorted. Thus, the Mulliken charge could be assigned quite accurately in these cases.
In the following section, we show that the weakened interface bonds represented preferred places for the dissociation of \Otwo molecules. 
Furthermore, interface regions between crystalline Si and the amorphous oxide of \SIrange{5}{7}{\angstrom} were indicated by TEM images~\cite{Diebold1999} and electron-energy-loss spectroscopy~\cite{Muller1999} (EELS) measurements.

\subsection{Oxygen migration}
Another important aspect of the surface oxidation process is the migration of incorporated oxygen.
First, we examined the possibilities of thermal diffusion of already adsorbed single O atoms by calculating energy barriers for the migration in Si bulk and from the surface oxide into deeper layers of the Si surface. Large diffusion barriers suggest that this mechanism is not important for the oxide growth.
In contrary, in later stages of the oxidation process, i.e. after a layer of about 5 to \SI{10}{\angstrom} of a-\SiOtwo formed on the surface, oxygen is incorporated by diffusion of \Otwo molecules through the amorphous oxide, as assumed within the Deal-Grove model.
We investigated this mechanism by AIMD simulations and confirmed the assumptions of Deal and Grove, namely, that oxygen molecules enter the oxide non-reactively and dissociate spontaneously when reaching the \interface interface.
Larger oxide models were used for these simulations since the interface of our DFTB obtained structure (see Fig.~\ref{SiO2_growth}) is still close to the surface and thus allows for the dissociative incorporation of molecular precursors as described above.

\subsubsection{Migration of adsorbed O atoms}

\begin{figure}[htbp]
	\includegraphics[width=\linewidth]{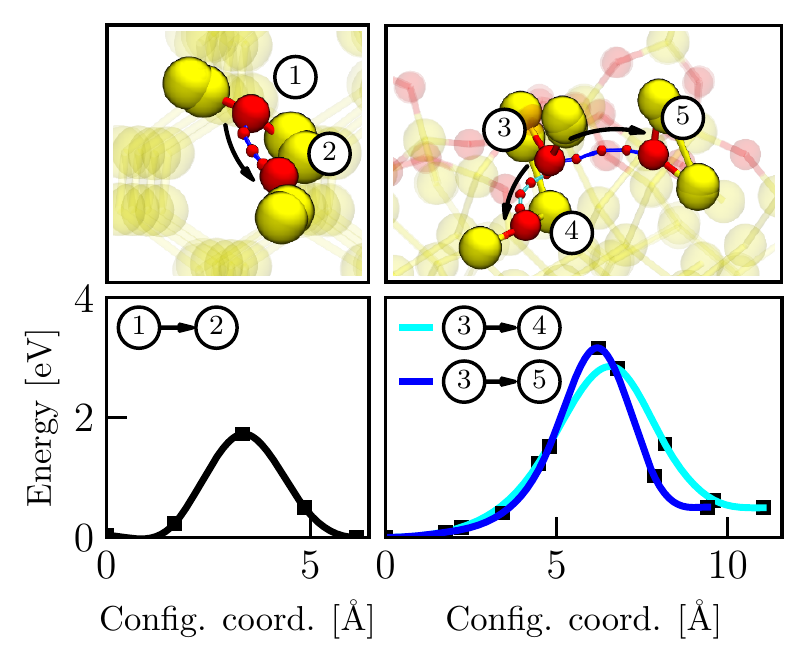}
    \caption{Energy barriers for thermal diffusion of O in Si bulk (left) and from a surface oxide layer (right) as obtained by NEB calculations. The trajectories of the diffusion are indicated by solid blue and cyan lines in the top panels and the resulting energy barriers are shown in the respective lower panels. Intermediate configurations along the trajectories are depicted as smaller O atoms. The results show that thermal diffusion of O from the oxidized surface layers is strongly inhibited compared to the diffusion in Si bulk and would only contribute marginally to the overall oxidation process that is almost exclusively driven by \Otwo diffusion through the oxide.}
    \label{thermal_diff}
\end{figure}

To obtain diffusion barriers for single O atom migration in a bulk Si system 
we first calculated the energy barrier for the migration of one single O atom in a Si bulk material. As verified within geometry optimizations, the energetically favored positions of an O atom incorporated in a Si crystal are the Si bond center sites~\cite{BOND1960}. Utilizing NEB calculations, we obtained the minimum energy path between two neighboring bond center sites. The energy barrier of \SI{1.72}{\electronvolt} in our pristine Si crystal agrees with other theoretical studies of O migration in Si~\cite{OinSi, OinSi2, OinSi3}.
With this result the experimentally derived energy barrier of \SI{2.5}{\electronvolt}~\cite{Tong1985} can be obtained by considering a coupled-barrier diffusion~\cite{CoupledBarrier}.
In order to derive an estimate for the rate of the process at $T=\SI{1000}{\kelvin}$ we employed the Arrhenius equation with an attempt frequency of \SI{6e12}{\per\second}~\cite{AttemptFrequency} and the \SI{1.72}{\electronvolt} barrier yielding \SI{9e4}{\per\second}.
Enlarged barriers were found for O migration from the surface oxide structures obtained by our AIMD simulations, see Fig.~\ref{thermal_diff}. We specifically investigated diffusion in the transition region between the crystalline and amorphous structures since it offered alternative positions for O atoms. Two examples of possible diffusion paths for one O atom were calculated and yielded energy barriers of at least \SI{2.8}{\electronvolt}, or a reaction rate of \SI{4e-4}{\per\second}. Keeping the experimental oxidation growth rate of about \SI{1}{\angstrom/\second}~\cite{Hoshino2000} for the initial stage of thermal oxidation of Si in mind, only a small contribution to the overall growth rate by thermal migration of single O atoms is expected. 
However, migration of O atoms is required for the dissociation of molecular precursors and thus still plays an important role in the oxidation process. The observation of slow oxygen migration in the amorphous surface layers together with an increased adsorption barrier 
explains the decrease of oxidation rate as the oxidation proceeds.
In this manner, the oxide growth rate decreases gradually as the dominant oxidation mechanism transitions from spontaneous surface reactions to precursor reactions and finally into the non-reactive \Otwo diffusion regime.

\subsubsection{\Otwo diffusion through the oxide}
Diffusion of \Otwo through thicker oxide layers during thermal oxidation as proposed by \cite{DealGrove} was well investigated in other studies~\cite{Bongiorno2005, Bongiorno2004, Bongiorno2004_2, Pasquarello1998, interface1988, AKIYAMA2005, Gusev1995, Rosencher1979, BAKOS2002, Hoshino2003, PEREZBUENO2000}. The diffusion barrier was found to be extremely sensitive to the local structural topology which is reflected by a wide spread of DFT calculated values from \SIrange{0.5}{2.8}{\electronvolt}~\cite{Bongiorno2004_2, BAKOS2002}.
Experimentally, values between 0.7 and \SI{1.6}{\electronvolt}~\cite{PEREZBUENO2000} were found for oxide films produced in various ways.
Our own NEB calculations of O$_2$ diffusion processes agree nicely with previous theoretical and experimental results and yielded barriers between 0.7 and \SI{2}{\electronvolt}. 
Hence, molecular stability as well as sufficient diffusibility of \Otwo in silicon dioxide can be safely assumed also within the framework of our calculations.

\begin{figure}[htbp]
	\centerline{\includegraphics[width=\linewidth]{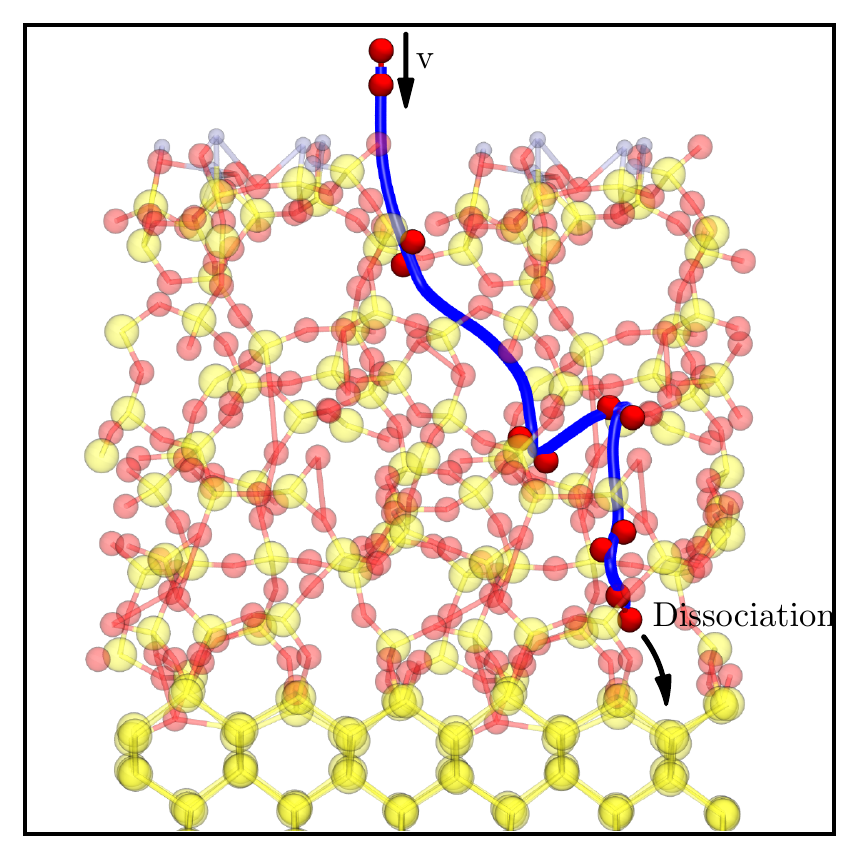}}
    \caption{Diffusion of \Otwo in a-\SiOtwo is enabled for thicker oxide layers. The non-reactive incorporation and migration through the oxide layer is shown by the results of an AIMD simulation with a simulation time of 3~ps. An initial velocity of 800~m/s was assigned to the \Otwo molecule. The trajectory of the center of mass of the molecule is depicted by a solid blue line. The molecule eventually reaches the \interface interface and dissociates, see Fig.~\ref{O2NEBinterface}.}
    \label{O2NEB}
\end{figure}

We observed the non-reactive incorporation of \Otwo molecules on a \SI{20}{\angstrom} thick oxide layer that was obtained by a melt and quench procedure~\cite{Jech}. As shown in Fig.~\ref{O2NEB}, a seamless migration through the oxide could be sampled in an AIMD simulation with a simulation time of \SI{1}{\pico\second}. 
Again, the molecules' initial velocity was set to \SI{1000}{\meter/\second}. 
Without interaction, the \Otwo entered the oxidized surface and migrated along a random path determined by deflections at the \SiOtwo structure. 
These results show that the molecular state of \Otwo represents a stable configuration in ultra-thin layers of a-SiO$_2$. As assumed within the Deal-Grove model~\cite{DealGrove}, the oxidation rate is now governed by the \Otwo diffusion rate. 

\subsubsection{\Otwo dissociation at the interface}
In the \Otwo diffusion regime, further oxidation of the Si substrate happens at the buried \interface interface upon dissociation of the \Otwo molecule~\cite{DealGrove}. From a macroscopic point of view, the O$_2$ migration yields an oxidation front that moves from the surface deeper into the substrate.
This process could be confirmed within our simulations by another set of AIMD calculations.

\begin{figure}[htbp]
	\centerline{\includegraphics[width=\linewidth]{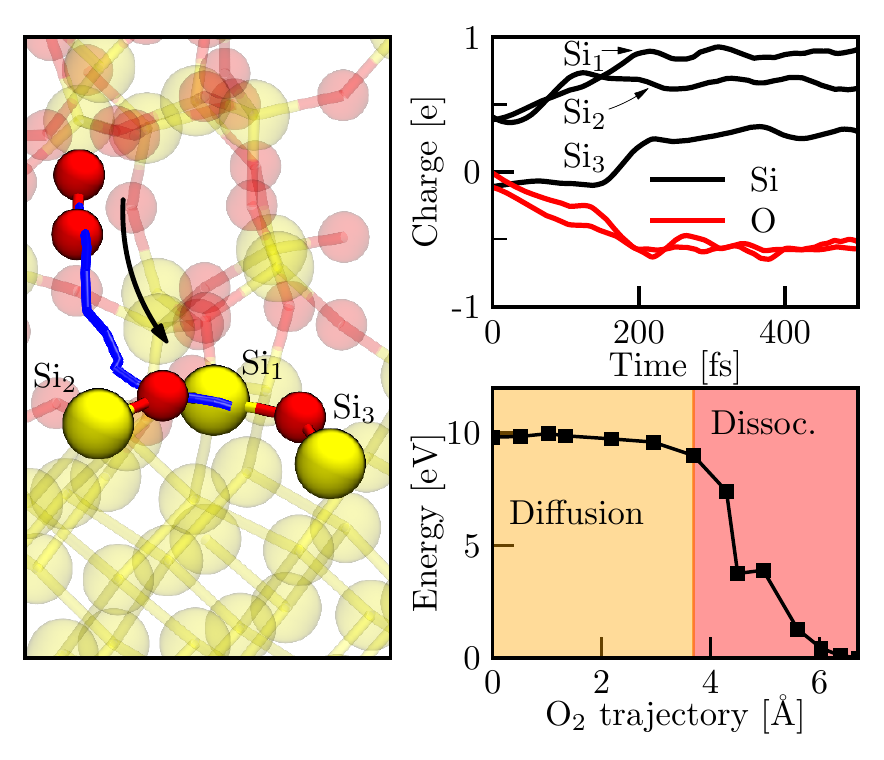}}
    \caption{\Otwo dissociation at the \interface interface. The left panel shows the trajectory of the center of mass of the \Otwo molecule (solid blue line) together with the initial (molecular) and final (dissociated) configurations of the NEB calculation. 
    The Mulliken charges for the spontaneous dissociation as captured within an AIMD simulation are depicted in the top right panel. The Mulliken analysis reveals another charge transfer of $-e$ that strongly resembles the dissociative reactions at the surface during the initial stage of oxidation, cf.~Fig.~\ref{charge_transfer}. The associated charges of the Si$_1$ and Si$_2$ atoms show an offset due to the implications of the interface, cf.~Fig.~\ref{q_vs_z}. The potential energy surface of this process is displayed in the lower right panel and indicates diffusion along a \SI{4}{\angstrom} long trajectory that is only inhibited by a low barrier of \SI{0.2}{\electronvolt}. In addition to the AIMD results, the NEB calculation provides evidence for a spontaneous dissociation as soon as the molecule reached the dissociation spot.}
    \label{O2NEBinterface}
\end{figure}

An \Otwo molecule was placed in the transition region between the amorphous oxide and its crystalline substrate. Due to lattice distortions, some Si atoms in the transition region possess strained bonds or even dangling bonds, cf. Fig.~\ref{q_vs_z}. These dangling bonds are known to affect the reliability of semiconductor devices and are therefore usually passivated with H after the oxidation~\cite{Jech, Hpassivation}. However, the simulations showed that these atoms were the preferred dissociation spots. The initial velocity of the \Otwo molecule was set to zero in order to verify the feasibility of a spontaneous reaction.
As shown in Fig.~\ref{O2NEBinterface}, the \Otwo molecule spontaneously dissociates via essentially the same charge transfer process as in the dissociative chemisorption at the Si surface, cf. Fig.~\ref{charge_transfer}. 
In addition, further evidence for a barrierless dissociation was obtained by another NEB calculation. Here, \Otwo was placed above the transition region in order to include a short diffusion path of \SI{4}{\angstrom} preceding the actual dissociation. The migration toward the strained bond was governed by a comparably low diffusion barrier of \SI{0.2}{\electronvolt}. In Fig.~\ref{O2NEBinterface}, the beginning of the dissociative process is indicated by a strong decrease of the potential energy surface. 
The large energy gain of \SI{9.4}{\electronvolt} could be justified by a complex reconfiguration at the interface in which multiple bonds with considerable binding energies break (O-O: \SI{2.1}{\electronvolt}; Si-Si: \SI{4.9}{\electronvolt}) and form (\hbox{Si-O}: \SI{6.5}{\electronvolt}).
Additionally, the molecular configuration contained the incorporation energy of the oxygen molecule in the oxide layer which was evaluated to be \SI{0.4}{\electronvolt} in bulk \SiOtwo by DFT calculations \cite{Stoneham2001}.
In this stage, the oxide growth rate is not limited by the supply of oxygen to the surface anymore but by the \Otwo diffusion rate through the oxide and is thus certainly much slower compared to the initial oxidation, as assumed within the Deal-Grove model~\cite{DealGrove}.
\section{Conclusions}
We modeled the oxidation of a Si(100) surface by means of ab-initio DFT and the closely related DFTB simulation techniques and obtained credible surface models of ultra thin native oxide layers on a Si substrate. 
For the first time, the highly complex thermal oxidation process could be consistently reproduced for ultra-thin SiO$_2$ layers ($<5$\AA) within simulations that reveal an oxidation scheme which combines all experimental observations that have been reported on this subject up to now such as surface reactions, adsorption into molecular precursor states and O$_2$ diffusion through the oxide~\cite{DealGrove, Bongiorno2005, Bongiorno2004, Bongiorno2004_2, Pasquarello1998, interface1988, AKIYAMA2005, Gusev1995, Rosencher1979, InitialAdsorp, InitialAdsorp2, InitialAdsorp3, Liao2006, expInitialAdsorp16, STM2020, Precursor, Precursor2, STM2020, Ferguson1999, Morgen1989, MB1999, Hoshino2001, MUR2001}.
By doing so, a complete picture of the whole process with its various mechanisms can be sketched.

Starting from the clean $p(2\times2)$  reconstructed Si surface we gradually introduced molecular oxygen into the system by adding \Otwo molecules above the surface. In this manner, we dynamically modeled the O$_2$ chemisorption until a \SI{3.5}{\angstrom} thick \SiOtwo layer formed above a \SI{5}{\angstrom} transition region.
Immediate amorphization from the onset of oxidation was indicated by a stochastic adsorption processes in which many barrierless adsorption trajectories with similar energy gains of around \SI{6}{\electronvolt} led to strongly varying final positions.
The top a-\SiOtwo layer showed many characteristics of bulk \SiOtwo such as geometric measures (bond angles and lengths) and density already for thicknesses of $\sim$\SI{3.5}{\angstrom}. Furthermore, it consisted of SiO$_4$ tetrahedrons just like the bulk oxide.  Thus, thermal oxidation of Si led to the immediate formation of a-SiO$_2$. The oxide layer was recognized by the oxidation states of the Si atoms, that is +4 for fully oxidized Si atoms incorporated in a SiO$_4$ tetrahedron.
The silicon dioxide was separated from the crystalline Si substrate by a \SI{5}{\angstrom} thick transition layer. This region was subjected to a significant inherent stress as indicated by strained bonds, compressed angles and an increased density.

In the initial oxidation stage, the adsorption of \Otwo was a spontaneous process and directly followed by the dissociation of the molecule upon which the O atoms moved into Si-Si bond center sites. Thus, this stage features the fastest oxidation rate that is only limited by the supply of oxygen.
Further oxidation of a already partly oxidized Si surface atom was found to require overcoming of an adsorption barrier.
Therefore, after oxidation of the first Si layer, the molecules were occasionally repelled and direct dissociation could only be observed for molecules with kinetic energies from the top one percent of the Maxwell-Boltzmann distribution.
However, the adsorption of \Otwo molecules into molecular precursor states that only dissociated after some picoseconds was commonly monitored in longer MD runs. The slower dissociation mechanism and the decreased adsorption probability led to an overall decrease of the oxide growth rate.
Finally, the oxide layer became thick enough to effectively block any surface reactions leading to the dissociation of the \Otwo molecule. Now the diffusion of \Otwo through the oxide to the \interface interface -- as assumed within the Deal-Grove model -- set in. 
As soon as the \Otwo molecule reached the Si/SiO$_2$ transition layer with a substantial amount of strained configurations, it again spontaneously dissociated via a charge transfer reaction that strongly resembled the surface reactions during the initial stage of oxidation.

Based on the qualitative behavior of the oxidation process, estimates for the transition into the diffusive Deal-Grove regime 
can be given.
Direct surface reactions dominate only during oxidation of the first and second Si layer (oxide thickness $\sim$\SI{4}{\angstrom}). Subsequently, in an intermediate stage, molecular precursor states provide for a slower oxidation until a layer of a-SiO$_2$ has formed above the $\sim$\SI{5}{\angstrom} transition region. Starting from \SI{10}{\angstrom} total oxide thickness, the surface should be inert to any surface reactions and thus allow for the diffusion of molecular oxygen within the a-\SiOtwo layer.
  
\section{Acknowledgments}
This project has received funding from the European
Union’s Horizon 2020 research and innovation programme
under grant agreement No. 871813, within the framework
of the project Modeling Unconventional Nanoscaled Device
FABrication (MUNDFAB). 
Furthermore, the financial support by the Austrian Federal Ministry for Digital and Economic Affairs and the National Foundation for Research, Technology and Development and from the Vienna Scientific Cluster (VSC) is gratefully acknowledged.

\bibliography{paper}

\end{document}